\documentclass[12pt]{article}

\usepackage{amsfonts}

\def\bR{{\mathbb R}} 

\begin{document}

\title{Strong Coupling Perturbation Theory in Quantum Mechanics}

\author{Marco Frasca \\
        Via Erasmo Gattamelata, 3,\\
        00176 Roma (Italy)}

\date{\today}

\maketitle

\abstract{
We present a full introduction to the recent devised perturbation theory
for strong coupling in quantum mechanics. In order to put the theory in a
proper historical perspective, the approach devised in quantum field theory is
rapidly presented, showing how it implies a kind of duality in perturbation
theory, from the start. The approach of renormalization group in perturbation theory
is then presented. This method permits
to resum secularities in perturbation theory and makes fully
algorithmical the resummation, transforming the perturbation calculations
in a step by step computational procedure. The general theorem on which is
founded a proper application of the strong coupling expansion, based on a result
in the quantum adiabatic theory, is then exposed. This theorem
gives the leading order of a strong coupling expansion. Then,
after the introduction of the principle of duality in perturbation theory
that puts in a proper context the quantum field theory method, the
resulting theory of the strong coupling expansion and the free picture
are presented. An algorithm for the computation of the perturbation series is
finally given. This approach has a lot of applications in fields as quantum
optics, condensed matter and so on, extending the original expectations of
the quantum field theory method. So, we give some examples of application for a class of two-level
systems that, in recent years, proved to be extremely important. One of the
most interesting concepts that can be obtained in this way is that of a
Quantum Amplifier (QAMP) that permits to obtain an amplification to the classical
level of the quantum fluctuations of the ground state of a single radiation mode.   
}

\newpage

\section{Introduction}

After the discovery of the quantum chromodynamics (QCD) and asymptotic freedom \cite{QCD},
it become increasingly important to treat in some way a theory having a non-perturbative
behavior. Since then, different approaches have been devised to recover the spectrum
of QCD at low energies but here we focus on a perturbative method that, although
did not prove to be useful to treat QCD problems, paved the way toward a strong
coupling expansion with a possible wider scope. This approach, the strong
coupling expansion, reached its best formulation in a paper by Bender and 
coworkers \cite{bend1,bend2,bend3} where, being applied to a $\lambda\phi^4$ quantum field theory,
it was proved that a lattice formulation could
give a manageable formulation but that the problem is moved on taking the
limit of zero lattice spacing on some very singular series for the relevant
quantities of the theory. Some resummation techniques were devised without
much success.

In quantum field theory it is customary to start with a path integral 
in the Euclidean space like for a $\lambda\phi^4$ theory 
(here and in the following $\hbar=c=1$)\cite{bend1} 
\begin{equation}
    Z[J]=\int D\phi\exp\left\{-\int dx\left[
	\frac{1}{2}(\partial\phi)^2+\frac{1}{2}m^2\phi^2+\frac{1}{4}\lambda\phi^4
	+J\phi
	\right]\right\}
\end{equation}
and to obtain a weak coupling expansion in $\lambda$ as
\begin{equation}
    Z[J]=
	\exp\left[-\frac{1}{4}\lambda\int dx\frac{\delta^4}{\delta J(x)^4}\right]
	\int D\phi\exp\left\{-\int dx\left[
	\frac{1}{2}(\partial\phi)^2+\frac{1}{2}m^2\phi^2+J\phi
	\right]\right\}
\end{equation}
where we have extract formally the quartic term from the path
integral and put a functional derivative for each power of the
field. Then, we are left with a gaussian integral that can be easily
computed and we have the sought expansion that can be cast in the form
\begin{equation}
    Z[J]=N\left\{1+\sum_{k=1}^\infty A_k[J]\lambda^k\right\}.
\end{equation}
At this stage one may ask what happens if we do the opposite operation,
that is, if we consider the kinetic term $\frac{1}{2}(\partial\phi)^2+\frac{1}{2}m^2\phi^2$
as a perturbation. This can be realized rewriting the path integral as
\begin{equation}
    Z[J]=\exp\left[-\frac{1}{2}\int\int dxdy
	\frac{\delta}{\delta J(x)}G^{-1}(x-y)\frac{\delta}{\delta J(y)}\right]
	\int D\phi\exp\left\{-\int dx\left[
	\frac{1}{4}\lambda\phi^4+J\phi
	\right]\right\}
\end{equation}
being $G^{-1}(x-y)=(-\partial^2+m^2)\delta(x-y)$
the inverse of the free Euclidean Green function.
It can be proven that, on a lattice being the theory highly singular,
the expansion takes the form \cite{bend1}
\begin{equation}
    Z[J]=N\left\{1+\sum_{k=1}^\infty B_k[J]\lambda^{-\frac{k}{2}}\right\}.
\end{equation}
where one can see a dependence on the inverse of the coupling constant $\lambda$.
We see that we have arbitrarily chosen different part of the Euclidean action
as a perturbation, a possibility offered by the freedom proper to this choice,
and, by doing that, we get two perturbation series having as development parameter
one the inverse of the other. This kind of ``duality'' can be found e.g. in
fluid mechanics with the Navier-Stokes equation where we can have different
perturbative regimes by taking as unperturbed part the Eulerian term or the
Navier-Stokes term \cite{fluid}. This regimes are characterized by large Reynolds
number and small Reynolds number respectively. This kind of duality in perturbation theory
can be indeed seen rather widely and appear an ubiquitous property
due to the freedom in the choice of what a perturbation is. The first
formulation in this sense in quantum mechanics appeared in Ref.\cite{fra0} following a series of
works where a strong perturbation theory in quantum mechanics has been formulated \cite{fra1,fra2,fra3,fra4}.
In this work it was shown that the adiabatic approximation, formulated as in \cite{most,most2},
is the leading order approximation of what can be called a dual Dyson series, as
the weak coupling expansion is the well-known Dyson series.

Being a dual series to the standard Dyson expansion it shares the same problems.
Particularly, one of the most relevant questions one has to face is that of 
secularities \cite{nayfeh,kc1,kc2}. A secularity is a polynomial contributions to the series
that increases without bound making the series itself useless. The name ``secularity''
is taken from celestial mechanics where firstly these terms appeared in perturbation
series with a secular timescale. Then, in order to have an useful tool to make
computations one has to devise a way to remove such singular terms. In the course
of time several approaches have been proposed to this aim \cite{nayfeh,kc1,kc2}.
The difficulties with these methods are essentially linked to the impossibility
to make them algorithmic in some way. But, recently a new approach has been
proposed \cite{cog1,cog2,kuni1,kuni2} that solved this problem making perturbation
computations straightforward to realize. This method relies on the renormalization
group techniques that aim to find the envelope of the computed perturbation series
to the desired order. This approach has been successfully applied in quantum
mechanics and to the dual Dyson series \cite{fra5,fra6,fra7,fra8,kuni3,egu}.

The existence of a strong coupling expansion can prove to be very important as it
gives the opportunity to study the solution of a differential equation in different
regimes in the parameter space. In turn, this means that new physics can be uncovered.
Although the strong coupling expansion has found several applications in different
fields as strong atom-laser interaction \cite{sal1,sal2},
quantum chaos \cite{fra9,fra9b,fra10} and quantum Zeno effect \cite{fp,fraz},
the workhorse for a lot of studies in fields as quantum optics, quantum computation
and condensed matter is the two-level system \cite{fra11,frac}. Then, to present a lot
of examples of application of the strong coupling expansion, we analyze some
examples of two-level Hamiltonians showing how relevant physics can be exploited
in this regime giving finally rise to the concept of a quantum amplifier (QAMP),
able to amplify the quantum fluctuations of a radiation mode in the ground state 
to the classical level \cite{fra12,fra13,fra13b,fra11}. 

This review is structured as follow. In Sec.\ref{sec1} we introduce the duality
principle in perturbation theory for the Schr\"odinger equation. In Sec.\ref{sec2}
the fundamental theorem of the strong coupling expansion to the leading order is
proved. In Sec.\ref{sec3} the problem of secularities is put forward as a general
problem in perturbation theory that is inherited by the strong coupling expansion.
In Sec.\ref{sec4} the renormalization group method to solve the secularity
problem is presented. In Sec.\ref{sec5} the approaches discussed above are merged
to formulate an algorithm for doing perturbation theory. In Sec.\ref{sec6} we 
apply the strong coupling expansion to a two-level model well-known in quantum
optics and finally, in sec.\ref{sec7} we show how the two-level model
discussed in Sec.\ref{sec6} can amplify the quantum fluctuation to the
classical level producing a classical field (QAMP). In Sec.\ref{sec8} the
conclusions are given.

\section{Duality in Perturbation Theory}
\label{sec1}

The starting point of our analysis is given by a quantum system with a Hamiltonian
\begin{equation}
    H=H_0+V.
\end{equation}
It is usually assumed that the dynamics for th $H_0$ part is known. Adding the
$V$ part can make the problem unmanageable unless one uses perturbation theory and
the $V$ part is smaller in some sense with respect to $H_0$. To account for this,
we introduce an arbitrary parameter $\lambda$ that we now consider small but we take
to be unity at the end of the computation. So, we write $H=H_0+\lambda V$ and aim
to solve the Schr\"odinger equation 
\begin{equation}
    (H_0+\lambda V)|\psi\rangle = i\frac{\partial|\psi\rangle}{\partial t}.
\end{equation}
When $\lambda$ is a small one does the transformation (interaction picture)
\begin{equation}
    |\psi\rangle = e^{-iH_0t}|\psi_I\rangle
\end{equation}
obtaining the equation to solve
\begin{equation}
    e^{iH_0t}\lambda Ve^{-iH_0t}|\psi_I\rangle = i\frac{\partial|\psi_I\rangle}{\partial t}.
\end{equation}
The solution of this equation, generally know as Dyson series, can be written as
\begin{equation}
    |\psi_I\rangle = {\cal T}\exp\left[-i\int_0^t dt'e^{iH_0t'}\lambda Ve^{-iH_0t'}\right]|\psi(0)\rangle
\end{equation}
being ${\cal T}$ the so called time ordering operator and $|\psi(0)\rangle$ the initial
wave function. This is just a formal writing for the small perturbation series
\begin{equation}
    |\psi_I\rangle = \left(I - i\int_0^t dt'e^{iH_0t'}\lambda Ve^{-iH_0t'}
	-\int_0^t dt'e^{iH_0t'}\lambda Ve^{-iH_0t'}\int_0^{t'} dt''e^{iH_0t''}\lambda Ve^{-iH_0t''}
	+\cdots\right)
\end{equation}
that we recognize as a power series, $|\psi_I\rangle=\sum_{n=0}^\infty \lambda^n|\phi_n(t)\rangle$,
that can have a meaning only for small values of $\lambda$.

We have done some assumptions to derive the Dyson series. We have assumed that $H_0$ is
the unperturbed part of the Hamiltonian whose dynamics is known and that $\lambda$
was a small parameter. Then, we have obtained a result that could be meaningful,
at least asymptotically. But we can relax both these assumptions in view of the fact
that the choice of an unperturbed part and a perturbation is totally arbitrary and
one may ask what would be the face of a series where the role of $H_0$ and $V$ is
interchanged. We can easily work out this exchange into the Dyson series obtaining
\begin{equation}
    |\psi_F\rangle = {\cal T}\exp\left[-i\int_0^t dt'e^{i\lambda Vt'}H_0e^{-i\lambda Vt'}\right]|\psi(0)\rangle.
\end{equation}
To understand what we have done we also transform the integration variable as $\tau = \lambda t'$
and we have
\begin{equation}
    |\psi_F\rangle = {\cal T}\exp\left[-i\frac{1}{\lambda}\int_0^{\lambda t} d\tau e^{iV\tau}H_0e^{-iV\tau}\right]
	|\psi(0)\rangle.
\end{equation}
that means the series
\begin{equation}
    |\psi_F\rangle = \left(I - i\frac{1}{\lambda}\int_0^{\lambda t} d\tau e^{iV\tau}H_0e^{-iV\tau}
	-\frac{1}{\lambda}\int_0^{\lambda t} d\tau e^{iV\tau}H_0e^{-iV\tau}
	\frac{1}{\lambda}\int_0^{\lambda\tau} d\tau' e^{iV\tau'}H_0e^{-iV\tau'}
	+\cdots\right).
\end{equation}
A meaning can be attached to this series only if we take the limit $\lambda\rightarrow\infty$ and
we have obtained a dual series with respect to the Dyson series having a development parameter inverse
to the latter. So, using a symmetry in the choice of what a perturbation is we were able to uncover
a new perturbation series, dual to the small perturbation series, useful in problems with a strong
perturbation. Finally, we can set $\lambda=1$ and we will have the Dyson series as
\begin{equation}
    |\psi_I\rangle = {\cal T}\exp\left[-i\int_0^t dt'e^{iH_0t'}Ve^{-iH_0t'}\right]|\psi(0)\rangle
\end{equation} 
after passing to the interaction picture with the unitary transformation $\exp(-iH_0t)$ and a dual
Dyson series
\begin{equation}
    |\psi_F\rangle = {\cal T}\exp\left[-i\int_0^t dt' e^{iVt'}H_0e^{-iVt'}\right]
	|\psi(0)\rangle.
\end{equation} 
passing to the free picture with the unitary transformation $\exp(-iVt)$. It is
easily realized that the two series will share the same kind of problems as
secularities, asymptotic convergence or divergence and so on. But, we are aware
that this does not diminish the usefulness of the perturbative approach.

\section{The Leading Order of the Strong Coupling Perturbation Theory}
\label{sec2}

The leading order of the dual Dyson series is quite straightforward to define if
the perturbation $V$ does not depends on time. Things are more involved otherwise.
Indeed, the dual Dyson series must be redefined if $V$ depends on time and we will
prove that a series, derived formally from the adiabatic approximation, due to
Mostafazadeh \cite{most}, is recovered in this case. Again we consider the problem
\begin{equation}
    (H_0+\lambda V(t))|\psi(t)\rangle = i\frac{\partial |\psi(t)\rangle}{\partial t} \label{eq:H}
\end{equation}
with $\lambda\rightarrow\infty$. To recover the dual Dyson series, we have to
determine the unitary operatore $U_F(t)$ such that
\begin{equation}
    \lambda V(t)U_F(t) = i\frac{\partial U_F(t)}{\partial t}
\end{equation}
with the same limit for $\lambda$. It is very easy to recognize here the starting
point of the proof of the adiabatic theorem in quantum mechanics\cite{messiah,most2} and
we are recovering a series, in the same framework, due to Mostafazadeh \cite{most,most2}.
This is due to the slowing down implied by the parameter $\lambda$ going to infinity.
Then, without recurring to any adiabatic hypothesis, we can write \cite{fra0}
\begin{equation}
    U_F(t) = \sum_n e^{i\gamma_n(t)} e^{-i\int_0^t v_n(t')dt'} |n;t\rangle\langle n;0|
\end{equation}
being $\gamma_n(t)$ the geometric part of the phase and $v_n(t)$ the dynamical part
such that $V(t)|n;t\rangle=v_n(t)|n;t\rangle$. We will show that this gives the
proper dual Dyson series. Higher order corrections can be written down and are
given in Refs.\cite{messiah,most,most2}. It is interesting to note that this theorem
has found an application in the studies of Zeno effect in quantum systems \cite{fp}.

Once the unitary operator $U_F(t)$ is known, we are able to pursue the
computation to the end. Indeed, we have to solve the Schr\"odinger equation
\begin{equation}
    H_F |\psi_F(t)\rangle = i\frac{\partial |\psi_F(t)\rangle}{\partial t} \label{eq:scF}
\end{equation}
being
\begin{equation}
    H_F = \sum_m\sum_n e^{i[\gamma_n(t)-\gamma_m(t)]}
	                   e^{-i\int_0^t[v_n(t)-v_m(t)]}
					   \langle m,t|H_0|n,t\rangle |m,0\rangle\langle n,0|
\end{equation}
the transformed Hamiltonian. We easily realize that the Hamiltonian $H_F$ can
be split in two parts as
\begin{eqnarray}
    H_F &=& \sum_n\langle n,t|H_0|n,t\rangle |n,0\rangle\langle n,0| \\ \nonumber
	    &+& \sum_{m\neq n} e^{i[\gamma_n(t)-\gamma_m(t)]}
	                   e^{-i\int_0^t[v_n(t')-v_m(t')]dt'}
					   \langle m,t|H_0|n,t\rangle |m,0\rangle\langle n,0|
\end{eqnarray}
and we are able to obtain the analogous equations of the interaction picture
for the probability amplitudes that apply in the case of the dual Dyson series.
In order to obtain this result we write the solution of eq.(\ref{eq:scF}) as
\begin{equation}
    |\psi_F(t)\rangle = \sum_n c_n(t)e^{-i\int_0^th_{0n}(t')dt'}|n,0\rangle 
\end{equation}
being
\begin{equation}
    h_{0n}(t)=\langle n,t|H_0|n,t\rangle.
\end{equation}
This gives the equations for the amplitudes
\begin{equation}
    i\dot c_n(t)=\sum_k e^{i[\gamma_k(t)-\gamma_n(t)]}
	e^{-i\int_0^t[\epsilon_k(t')-\epsilon_n(t')]dt'}
	\langle n,t|H_0|k,t\rangle c_k(t)
\end{equation}
with $\epsilon_n(t)=h_{0n}(t)+v_n(t)$.
These equations are similar to the amplitude equations in interaction picture, normally
found in textbooks.

Some considerations are in order at this point. By the adiabatic theorem we
can only apply the above approach for a discrete spectrum on the perturbation.
This in turn implies that, if the spectrum of the perturbation is continuous as
in the coordinate space, then a lattice regularization is needed, This takes us
back to Bender et al. approach \cite{bend1,bend2,bend3} 
with all the difficulties this means by taking
the limit of the lattice spacing going to zero.
 
\section{The Secularity Problem in Perturbation Theory}
\label{sec3}

A perturbation series is plagued by secularities when polynomial terms in time
appears in it. These terms have the property of being not bounded for large
time making the series generally useless.

In order to have an idea of what really happens in these situations, let us
consider the well known quantum mechanical problem
\begin{equation}
    H = \frac{\Delta}{2}\sigma_3 + g \sigma_1 \cos(\omega t) \label{eq:hs}
\end{equation}
representing a two level atom driven by an oscillating field with frequency $\omega$.
$\sigma_1$,$\sigma_3$ are Pauli matrices, $g$ is the coupling constant and $\Delta$
is the separation between the atom levels. This problem has a large body of
literature due to its vast field of applications (see e.g. Ref.\cite{fra8,fra11}).

We can apply to this problem all the machinery devised in the preceding section
for the strong coupling expansion. So, one has
\begin{equation}
    U_F(t)=e^{-i\frac{g}{\omega}\sin(\omega t)}|+\rangle\langle +|+
	e^{i\frac{g}{\omega}\sin(\omega t)}|-\rangle\langle -|
\end{equation}
being $\sigma_1|\pm\rangle = \pm|\pm\rangle$. By using the explicit expression for
the states $|\pm\rangle$ one can prove that $U_F(t)=e^{-i\sigma_1\frac{g}{\omega}\sin(\omega t)}$
as it should be for the free picture. It is very easy to obtain
\begin{equation}
    H_F=\frac{\Delta}{2}\left[e^{-i\frac{2g}{\omega}\sin(\omega t)}|-\rangle\langle +|+
	e^{i\frac{2g}{\omega}\sin(\omega t)}|+\rangle\langle -|\right].
\end{equation}   
Looking for a solution in the form $|\psi_F(t)\rangle = c_+(t)|+\rangle+c_-(t)|-\rangle$
we obtain the equations for the amplitudes
\begin{eqnarray}
    i\dot c_+(t)&=&\frac{\Delta}{2}e^{i\frac{2g}{\omega}\sin(\omega t)}c_-(t) \\ \nonumber
	i\dot c_-(t)&=&\frac{\Delta}{2}e^{-i\frac{2g}{\omega}\sin(\omega t)}c_+(t)
\end{eqnarray}
that give rise to the perturbation series till first order
\begin{eqnarray}
    c_+(t)&=&c_+(0)-i\frac{\Delta}{2}J_0(z)c_-(0)t-\frac{\Delta}{2}
	\sum_{n\neq 0}J_n(z)\frac{e^{in\omega t}-1}{n\omega}c_-(0)+\cdots \\ \nonumber
	c_-(t)&=&c_-(0)-i\frac{\Delta}{2}J_0(z)c_+(0)t+\frac{\Delta}{2}
	\sum_{n\neq 0}J_n(z)\frac{e^{-in\omega t}-1}{n\omega}c_+(0)+\cdots
\end{eqnarray}
where use has been made of the relation $e^{iz\sin(\omega t)}=\sum_{n=-\infty}^\infty J_n(z)e^{in\omega t}$
with $J_n(z)$ the n-th Bessel function of integer order and $z=\frac{2g}{\omega}$ in our case.
We see immediately that the perturbation series is plagued with secularities and so
is useless at this stage. We have to understand what is going on here by properly
resum such terms. In this way, we will discover here a physical effect, i.e. Rabi
oscillations between the states $|\pm\rangle$. The resummation technique to
accomplish our task is described in the next section where we will complete our
computation.

\section{Renormalization Group Method for the Resummation of Secular Terms}
\label{sec4}

The method of renormalization group to resum secularities in a perturbation
series was firstly proposed in \cite{cog1,cog2}. Here we present an elegant reformulation
obtained by the mathematical theory of envelopes by Kunihiro \cite{kuni1,kuni2}.

Kunihiro approach can be described as follows. Let us consider the following equation
\begin{equation}
    \dot x(t)=f(x(t),t) \label{eq:xf}
\end{equation}
being x(t) a vector in $\bR^n$. The initial condition is given by $x(t_0)=X(t_0)$.
At this stage we assume $X(t_0)$ not yet specified. We write the solution of
this equation as $x(t;t_0,X(t_0))$ that is exact. If we change $t_0$ to $t_0'$
we are able to determine $X(t_0)$ by assuming that the solution should not change
\begin{equation}
    x(t;t_0,X(t_0)) = x(t;t_0',X(t_0'))
\end{equation} 
that in the limit $t_0\rightarrow t_0'$ becomes
\begin{equation}
     \frac{dx}{dt_0}=\frac{\partial x}{\partial t_0}+
	 \frac{\partial x}{\partial X}\frac{\partial X}{\partial t_0}=0
\end{equation}
giving the evolution equation or flow equation of the initial value $X(t_0)$.
We recognize here a renormalization group equation and this gives the name to the
method. 

Till now, all our equations are exact and no perturbation theory entered in any part
of our argument. But, except for a few cases, the solution $x(t;t_0,X(t_0))$ is
only known perturbatively and such a solution are generally valid only locally,
i.e. for $t\sim t_0$ and $t\sim t_0'$ and a more restrictive request should be
demanded to our renormalization group equation
\begin{equation}
     \left.\frac{dx}{dt_0}\right|_{t_0=t}=
	 \left.\frac{\partial x}{\partial t_0}\right|_{t_0=t}+
	 \left.\frac{\partial x}{\partial X}\frac{\partial X}{\partial t_0}\right|_{t_0=t}=0. \label{eq:rg}
\end{equation}
But this equation can be interpreted by the mathematical theory of envelopes \cite{kuni1}.
Indeed, varying $t_0$ we have that $x(t;t_0,X(t_0))$ is a family of curves
with $t_0$ being a characterizing parameter. Then, eq.(\ref{eq:rg}) becomes
an equation to compute the envelope of such a family of curves. Such an envelope
is given by $x(t;t_0=t)=X(t)$, the initial condition. It can be proven that
$X(t)$ satisfies the equation (\ref{eq:xf}) in a global domain up to the order with which
$x(t;t_0)$ satisfies it locally for $t\sim t_0$. This gives the condition for
the computation of the envelope
\begin{equation}
    \left.\frac{dx}{dt_0}\right|_{t_0=t}=0 \label{eq:ee}
\end{equation} 

The Kunihiro method is very
effective to build resummed perturbation series, eliminating the seculairities
that appear to plague them. Besides, it permits to transform a perturbation
computation in an algorithm straightforward to apply as we are going to see in the next
section.

\section{An Algorithm for Doing Perturbation Theory in Quantum Mechanics}
\label{sec5}

In order to exploit what we mean by an algorithmic computation of a perturbation
series, we come back to the example given in sec.\ref{sec3}. All we have to do,
as our first step, is to recompute the perturbation series at a generic initial
time $t_0$ and to assume generic initial conditions. This yields
\begin{eqnarray}
    c_+(t)&=&\tilde c_+(t_0)-i\frac{\Delta}{2}J_0(z)\tilde c_-(t_0)(t-t_0)-\frac{\Delta}{2}
	\sum_{n\neq 0}J_n(z)\frac{e^{in\omega t}-e^{in\omega t_0}}{n\omega}\tilde c_-(t_0)+\cdots \\ \nonumber
	c_-(t)&=&\tilde c_-(t_0)-i\frac{\Delta}{2}J_0(z)\tilde c_+(t_0)(t-t_0)+\frac{\Delta}{2}
	\sum_{n\neq 0}J_n(z)\frac{e^{-in\omega t}-e^{-in\omega t_0}}{n\omega}\tilde c_+(t_0)+\cdots.
\end{eqnarray}
We realize easily that the envelope could not be computed with this series as is.
What we need here is to dress all the phases in the exponentials. This gives
\begin{eqnarray}
    c_+(t)&=&\tilde c_+(t_0)-i\frac{\Delta}{2}J_0(z)\tilde c_-(t_0)(t-t_0)-\frac{\Delta}{2}
	\sum_{n\neq 0}J_n(z)\frac{e^{in\omega t}-e^{-in\omega\phi(t_0)}}{n\omega}\tilde c_-(t_0)+\cdots \\ \nonumber
	c_-(t)&=&\tilde c_-(t_0)-i\frac{\Delta}{2}J_0(z)\tilde c_+(t_0)(t-t_0)+\frac{\Delta}{2}
	\sum_{n\neq 0}J_n(z)\frac{e^{-in\omega t}-e^{in\omega\phi(t_0)}}{n\omega}\tilde c_+(t_0)+\cdots.
\end{eqnarray}
where we have introduced a renormalizable phase $\phi(t_0)=-t_0$. This is strictly
linked to the property of quantum systems to have a freedom in the choice of the
initial phase. At this point we use the renormalization group or envelope equation
(\ref{eq:ee}) giving
\begin{eqnarray}
     \frac{\partial\tilde c_+(t)}{\partial t}+i\frac{\Delta}{2}J_0(z)\tilde c_-(t)+\ldots &=& 0 \\ \nonumber
	 \frac{\partial\tilde c_-(t)}{\partial t}+i\frac{\Delta}{2}J_0(z)\tilde c_+(t)+\ldots &=& 0 \\ \nonumber
	 \frac{\partial\phi(t)}{\partial t}+\ldots&=&0
\end{eqnarray}
and the perturbation solution is then given by
\begin{eqnarray}
    c_+(t)&=&\tilde c_+(t)-\frac{\Delta}{2}
	\sum_{n\neq 0}J_n(z)\frac{e^{in\omega t}-1}{n\omega}\tilde c_-(t)+\cdots \\ \nonumber
    c_-(t)&=&\tilde c_-(t)+\frac{\Delta}{2}
	\sum_{n\neq 0}J_n(z)\frac{e^{-in\omega t}-1}{n\omega}\tilde c_+(t)+\cdots.
\end{eqnarray}
completely solving our problem, till first order, and having all secular terms
properly removed in an algorithmic and simple way.

We can finally exploit fully our algorithm for doing perturbation theory in
presence of secular terms in the strong coupling regime for quantum mechanics. 
The rules for computing the unitary evolution operator, at any desired
order, are the following \cite{fra8}:

\begin{enumerate}
\item Consider the following unitary transformation on the equation (\ref{eq:H}) \cite{fra0}
\begin{equation}
    U_F(t) = \sum_n e^{i\gamma_n(t)} e^{-i\int_0^t v_n(t')dt'} |n;t\rangle\langle n;0| 
\end{equation}
with the eigenstates of the perturbation $|n;t\rangle$. This gives the transformed Hamiltonian
\begin{equation}
     H_F(t) = U_F^\dagger(t)H_0U_F(t).
\end{equation}
The dual Dyson series is computed by \cite{fra1,fra2,fra3,fra4}
\begin{equation}
     S_D(t,t_0) = {\cal T}\exp\left[-i\epsilon\int_{t_0}^tH_F(t')dt'\right]
\end{equation}
being as usual ${\cal T}$ the time ordering operator and an ordering parameter $\epsilon$
has been introduced that will be taken unity at the end of computation. It is fundamental for our
argument that the computation of this series is performed at a different starting point $t_0$.
\item Assume, at the start, that the time evolution operator has the form
\begin{equation}
    U(t,t_0) = U_F(t)S_D(t,t_0)U_R(t_0)
\end{equation}
where $U_R(t_0)$ is a ``renormalizable'' part of the unitary evolution.
\item At the given order one gets $S_D(t,t_0)$ as
\begin{equation}
    S_D(t,t_0) = I - i\epsilon f_1(t,t_0) -\epsilon^2 f_2(t,t_0)+\ldots
\end{equation}
and, at this stage, if some oscillating functions in $t_0$ appear like $e^{-i\omega t_0}$
then introduce the phase $\phi(t_0) = -t_0$ as a ``renormalizable'' parameter
rewriting it as $e^{i\omega\phi(t_0)}$. The secularities must be left untouched.
\item Eliminate the dependence on $t_0$ by requiring\cite{kuni1,kuni2}
\begin{equation}
     \left.\frac{dU(t,t_0)}{dt_0}\right|_{t_0=t} = 0 \label{eq:rgcond}
\end{equation}
and one obtains the renormalization group equations
\begin{eqnarray}
    \frac{dU_R(t)}{dt} &=& \epsilon g_1 U_R(t) + \epsilon^2 g_2 U_R(t) + O(\epsilon^3) \\ \nonumber
	\frac{d\phi(t)}{dt} &=& \epsilon \phi_1 \phi(t) + \epsilon^2 \phi_2 \phi(t) + O(\epsilon^3) 
\end{eqnarray}
where, at some stage, to obtain such equations at the second order, we have to use
their expressions at the first order as, to compute their form at order $n$-th one
have to use these equations at the order $(n-1)$-th, into the condition (\ref{eq:rgcond}).
This is a step toward the computation of the envelope of the perturbation series as said 
in sec.\ref{sec4}.
\item Finally, the renormalization equations should be solved and substituted into the
equation
\begin{equation}
     \left.U(t,t_0)\right|_{t_0=t} \label{eq:u}
\end{equation}
giving the solution, i.e. the envelope, we were looking for without secularities at the order
we made the computation.
\end{enumerate}

Once the unitary evolution is known, we can easily compute the wave function,
given the initial condition, recovering the case we have shown of the driven
two-level system.

We are going to see this approach at work in the next sections.

\section{Two-Level Systems and the Strong Coupling Perturbation Theory}
\label{sec6}

In quantum optics the interaction between a single radiation mode and a two-level
atom proves to be a paradigm for most of applications \cite{qo1,qo2,qo3}. The
Hamiltonian is 
\begin{equation}
    H = \frac{\Delta}{2}\sigma_z + \omega a^\dagger a + g\sigma_x (a^\dagger + a) \label{eq:hq}
\end{equation}
that differs from Hamiltonian (\ref{eq:hs}) by having a fully quantized radiation
field of frequency $\omega$, rather than a classical field, whose creation and
annihilation operator are $a^\dagger$ and $a$.

The standard approach \cite{fra11} to this problem is given by doing the unitary transformation
\begin{equation}
     U_I = e^{-i\frac{\Delta}{2}\sigma_z t}e^{-i\omega a^\dagger a t}
\end{equation}
and then the rotating wave approximation is applied keeping only the near resonant
terms ($\Delta\approx\omega$). This reduces our model to the well-known 
Jaynes-Cummings Hamiltonian \cite{qo1,qo2}
\begin{equation}
     H_{JC} = \frac{\delta}{2}\sigma_z + g(a\sigma_++a^\dagger\sigma_-)
\end{equation}
being now $\delta$ the detuning between the frequency of the radiation field $\omega$
and the separation between the levels of the atom $\Delta$. This aspect is well
known having a large body of literature since its inception and being a foundational
matter for quantum optics.

Instead, here our aim is to realize a complete study of the Hamiltonian (\ref{eq:hq})
from the point of view of our strong coupling approach \cite{fra16a,fra16,fujii1,fujii2}. For our
aims we have to compute $U_F(t)$ and this is done by solving the eigenvalues problem
\begin{equation}
     [\omega a^\dagger a + g\sigma_x(a^\dagger + a)]|[n;\lambda]\rangle = E_{n,\lambda}|[n;\lambda]\rangle
\end{equation}
whose solution is given by
\begin{equation}
    |[n;\lambda]\rangle=e^{\frac{g}{\omega}\lambda(a-a^\dagger)}|n\rangle|\lambda\rangle
\end{equation}
with $\sigma_x|\lambda\rangle=\lambda|\lambda\rangle$, $\lambda=\pm 1$ and the
eigenvalues, independent on $\lambda$ being $E_n=n\omega-\frac{g^2}{\omega}$. Then,
it is straightforward to write the unitary evolution operator as
\begin{equation}
   U_{F0}(t)=\sum_{n,\lambda}e^{-iE_nt}|[n;\lambda]\rangle\langle[n;\lambda]|
	|\lambda\rangle\langle\lambda| \label{eq:uf}
\end{equation}
that gives rise to the Hamiltonian
\begin{equation}
   H_F = U_{F0}^\dagger(t)\frac{\Delta}{2}\sigma_z U_{F0}(t)
\end{equation}
It is easily realized that it can be rewritten in the form \cite{fra16a,fra16,fujii1,fujii2}
\begin{equation}
    H_F=H_0'+H_1
\end{equation}
being
\begin{equation}
    H'_0=\frac{\Delta}{2}\sum_n e^{-\frac{2g^2}{\omega^2}}
	L_n\left(\frac{4g^2}{\omega^2}\right)
	\left[
	|[n;1]\rangle\langle[n;-1]||1\rangle\langle -1|+
	|[n;-1]\rangle\langle[n;1]||-1\rangle\langle 1|
	\right] \label{eq:h10}
\end{equation}
being $L_n$ the n-th Laguerre polynomial \cite{grad} and
\begin{eqnarray}
    H_1&=&\frac{\Delta}{2}\sum_{m,n,m\neq n}e^{-i(n-m)\omega t}
	\left[
	\langle n|e^{-\frac{2g}{\omega}(a-a^\dagger)}|m\rangle
	|[n;1]\rangle\langle[m;-1]||1\rangle\langle -1|+
	\right.\\ \nonumber
	& &\left.\langle n|e^{\frac{2g}{\omega}(a-a^\dagger)}|m\rangle
	|[n;-1]\rangle\langle[m;1]||-1\rangle\langle 1|
	\right].
\end{eqnarray}
At this point we can iterate the procedure by diagonalizing the Hamiltonian (\ref{eq:h10}).
The eigenstates are
\begin{equation}
    |\psi_n;\sigma\rangle=\frac{1}{\sqrt{2}}
	\left[
	\sigma|[n;1]\rangle|1\rangle+
	|[n;-1]\rangle|-1\rangle
	\right]
\end{equation} 
and the eigenvalues are
\begin{equation}
    \tilde E_{n,\sigma}=\sigma\frac{\Delta}{2}e^{-\frac{2g^2}{\omega^2}}
	L_n\left(\frac{4g^2}{\omega^2}\right)
\end{equation} 
being $\sigma=\pm 1$. So, we can write the unitary transformation
\begin{equation}
    U_{F1}(t)=\sum_{n,\sigma} e^{-i\tilde E_{n,\sigma}t}|\psi_n;\sigma\rangle\langle\psi_n;\sigma| 
\end{equation}
and get the transformed Hamiltonian
\begin{equation}
    H_1'=U_{F1}^{\dagger}(t)H_1U_{F1}(t)
\end{equation}
that is
\begin{equation}
    H_1'=\frac{\Delta}{2}\sum_{m,n,m\neq n}\sum_{\sigma_1,\sigma_2}
	{\cal R}_{mn,\sigma_1\sigma_2} e^{-i[(n-m)\omega-(\tilde E_{n,\sigma_1}-\tilde E_{m,\sigma_2})]t}
	|\psi_n;\sigma_1\rangle\langle\psi_m;\sigma_2|
\end{equation}
being
\begin{equation}
    {\cal R}_{mn,\sigma_1\sigma_2}=\frac{1}{2}\left[
	\langle n|e^{-\frac{2g}{\omega}(a-a^\dagger)}|m\rangle\sigma_1 +
	\langle n|e^{\frac{2g}{\omega}(a-a^\dagger)}|m\rangle\sigma_2
	\right].
\end{equation}
So, we have accomplished the unitary transformation
\begin{equation}
    U_F(t)=U_{F0}(t)U_{F1}(t)
\end{equation}
and we are left with the Schr\"odinger equation for our aims
\begin{equation}
    H_1'S_D(t,t_0) = i\frac{\partial S_D(t,t_0)}{\partial t}
\end{equation}
that we solve by the perturbation theory obtaining the strong coupling
expansion for this problem. As said in the formulation of the algorithm
in Sec.\ref{sec5}, we assume a solution in the form
\begin{equation}
     U(t,t_0) = U_F(t)S_D(t,t_0)U_R(t_0)
\end{equation}
being $U_R(t_0)$ a renormalizable part of the unitary evolution.

In order to obtain the sought series we need to understand where resonances
occur, that is where the condition $(n-m)\omega-(\tilde E_{n,\sigma_1}-\tilde E_{m,\sigma_2})=0$
is met. This happens for $n\neq m$ and for the two other conditions $\sigma_1=\sigma_2$
(intraband resonance) or $\sigma_1\neq\sigma_2$ (interband resonance). Then,
as required by our algorithm, we compute the dual Dyson series at an initial
time $t_0$ obtaining
\begin{eqnarray}
    S_D(t,t_0)&=&I - i\frac{\Delta}{2}\left[
	\sum_{intraband}{\cal R}_{mn,\sigma_1\sigma_2}|\psi_n;\sigma_1\rangle\langle\psi_m;\sigma_2|(t-t_0)\right. \\ \nonumber
    &+&\sum_{interband}{\cal R}_{mn,\sigma_1\sigma_2}|\psi_n;\sigma_1\rangle\langle\psi_m;\sigma_2|(t-t_0)\\ \nonumber
	&+&\sum_{\stackrel{m,n,\sigma_1,\sigma_2}{out~of~resonance}}
	{\cal R}_{mn,\sigma_1\sigma_2}\times \\ \nonumber
	& &\left.\frac{e^{-i[(n-m)\omega-(\tilde E_{n,\sigma_1}-\tilde E_{m,\sigma_2})]t}-
	e^{-i[(n-m)\omega-(\tilde E_{n,\sigma_1}-\tilde E_{m,\sigma_2})]t_0}}
	{-i[(n-m)\omega-(\tilde E_{n,\sigma_1}-\tilde E_{m,\sigma_2})]}	
	|\psi_n;\sigma_1\rangle\langle\psi_m;\sigma_2|
	\right]+\ldots.
\end{eqnarray}
The next step is to introduce the phase $\phi(t_0)=-t_0$ into the exponentials changing the series into
\begin{eqnarray}
    S_D(t,t_0)&=&I - i\frac{\Delta}{2}\left[
	\sum_{intraband}{\cal R}_{mn,\sigma_1\sigma_2}|\psi_n;\sigma_1\rangle\langle\psi_m;\sigma_2|(t-t_0)\right. \\ \nonumber
    &+&\sum_{interband}{\cal R}_{mn,\sigma_1\sigma_2}|\psi_n;\sigma_1\rangle\langle\psi_m;\sigma_2|(t-t_0)\\ \nonumber
	&+&\sum_{\stackrel{m,n,\sigma_1,\sigma_2}{out~of~resonance}}
	{\cal R}_{mn,\sigma_1\sigma_2}\times \\ \nonumber
	& &\left.\frac{e^{-i[(n-m)\omega-(\tilde E_{n,\sigma_1}-\tilde E_{m,\sigma_2})]t}-
	e^{i[(n-m)\omega-(\tilde E_{n,\sigma_1}-\tilde E_{m,\sigma_2})]\phi(t_0)}}
	{-i[(n-m)\omega-(\tilde E_{n,\sigma_1}-\tilde E_{m,\sigma_2})]}	
	|\psi_n;\sigma_1\rangle\langle\psi_m;\sigma_2|
	\right]+\ldots.
\end{eqnarray}
Finally, we can compute the envelope of $U(t,t_0)$ obtaining the renormalization group equations
\begin{eqnarray}
    \frac{dU_R(t)}{dt}&=&-i\frac{\Delta}{2}\left[
	\sum_{intraband}{\cal R}_{mn,\sigma_1\sigma_2}|\psi_n;\sigma_1\rangle\langle\psi_m;\sigma_2|\right. \\ \nonumber
    & &+\left.\sum_{interband}{\cal R}_{mn,\sigma_1\sigma_2}
	|\psi_n;\sigma_1\rangle\langle\psi_m;\sigma_2|\right]U_R(t)+\ldots \\ \nonumber
	\frac{d\phi(t)}{dt}&+&\ldots=0.
\end{eqnarray}
Then, computing $U(t,t_0)|_{t_0=t}$, we get the series
\begin{eqnarray}
    U(t)&=&U_F(t)\left[
	I + \frac{\Delta}{2}\sum_{\stackrel{m,n,\sigma_1,\sigma_2}{out~of~resonance}}
	{\cal R}_{mn,\sigma_1\sigma_2}\times\right. \\ \nonumber
	& &\left.\frac{e^{-i[(n-m)\omega-(\tilde E_{n,\sigma_1}-\tilde E_{m,\sigma_2})]t}-1}
	{(n-m)\omega-(\tilde E_{n,\sigma_1}-\tilde E_{m,\sigma_2})}	
	|\psi_n;\sigma_1\rangle\langle\psi_m;\sigma_2|+\ldots\right]\times \\ \nonumber
	& &U_R(t).
\end{eqnarray}

The relevant result is that, by the renormalization group method, we have
resummed the perturbation series obtaining the unitary evolution, $U_R(t)$,
proper to Rabi oscillations as it should be \cite{fra16a,fra16,fujii1,fujii2} plus a first
order correction. Rabi oscillations in the strong coupling regime, as described
here, have been recently observed in Josephson junctions \cite{naka}.

\section{An Application: The Quantum Amplifier (QAMP)}
\label{sec7}

The next step is to generalize the model of sec.\ref{sec6} to $N$ two-level
atoms. We will find a new physical effect that can be seen as a quantum amplification
of the vacuum fluctuations, that is, we realize a quantum amplifier (QAMP). For
our aims, it is very easy to generalize the Hamiltonian (\ref{eq:hq}) as 
\begin{equation}
    H_N = \frac{\Delta}{2}\sum_{i=1}^N\sigma_{zi} + 
	\omega a^\dagger a + g\sum_{i=1}^N\sigma_{xi} (a^\dagger + a). \label{eq:hqN}
\end{equation}
Now, we introduce the analogous of angular momentum operators as
\begin{eqnarray}
    S_x &=&  \frac{1}{2}\sum_{i=1}^N\sigma_{xi} \\ \nonumber
	S_y &=&  \frac{1}{2}\sum_{i=1}^N\sigma_{yi} \\ \nonumber
	S_z &=&  \frac{1}{2}\sum_{i=1}^N\sigma_{zi} \\ \nonumber
	S^2 &=& S_x^2+S_y^2+S_z^2
\end{eqnarray} 
with the well-known commutation relations $[S,S_i]=0$ and $[S_i,S_j]=i\epsilon_{ijk}S_k$,
with the index $i,j,k$ that can take the values $x,y,z$. Now, we have that,
depending on $N$ being even we can have a zero momentum state, otherwise \cite{dicke}
\begin{eqnarray}
      |S_x| <= &S& <= \frac{N}{2} \\ \nonumber
	  -\frac{N}{2}<= &S_x& <= \frac{N}{2}
\end{eqnarray}
giving the Dicke states $|S,S_x\rangle$.

At this stage, we can iterate the procedure in sec.\ref{sec6} by diagonalizing the
Hamiltonian
\begin{equation}
    H_F = \omega a^\dagger a + 2gS_x (a^\dagger + a)
\end{equation}
with the eigenstates
\begin{equation}
    |[n;S,S_x]\rangle = e^{\frac{2g}{\omega}S_x(a-a^\dagger)}|n\rangle|S,S_x\rangle
\end{equation}
and eigenvalues
\begin{equation}
    E_{n,S_x}=\left[n-\frac{4g^2S_x^2}{\omega^2}\right]\omega
\end{equation}
and this time we have no degeneracy with respect to the Dicke states as happened
for the single two-level atom. Then, it is straightforward to write down
the unitary transformation as
\begin{equation}
    U_F(t) = \sum_n\sum_{S,S_x}e^{-iE_{n,S_x}t}|[n;S,S_x]\rangle\langle[n;S,S_x]|.
\end{equation}
Already at this stage we can have quantum amplification. Indeed, let us take
as initial state $|\psi(0)\rangle = |0\rangle|\frac{N}{2},\frac{N}{2}\rangle$,
that is, we have the radiation field in the ground state and the maximal Dicke
state. In the ground state, the radiation field has vacuum fluctuations as it
is well-known.  This gives
\begin{equation}
    |\psi(t)\rangle = U_F(t)|\psi(0)\rangle = \sum_n e^{-iE_{n,\frac{N}{2}}t}
	|[n;\frac{N}{2},\frac{N}{2}]\rangle
	e^{-\frac{\alpha_N^2}{2}}\frac{\alpha_N^n}{\sqrt{n!}}
\end{equation}
that is nothing else that the solution of Ref.\cite{fra12,fra13,fra13b,fra11}, that is, a
coherent state with a parameter increasing for large N. At this stage, we can
take two different thermodynamic limits. The first one is given by statistical
mechanics, i.e. $N\rightarrow\infty$, $V\rightarrow\infty$ and $\frac{N}{V}=const$,
being $V$ the volume that contains the radiation mode (e.g. a cavity). The second
one is given just by the limit $N\rightarrow\infty$ keeping fixed the volume.
In the former case, we observe that $g\propto\frac{1}{V^\frac{1}{2}}$ and so,
the thermodynamic limit gives a classical radiation state in the thermodynamic
limit. In the latter case, the result is similar but we have the parameter
of the coherent state increasing as $N$, i.e. faster. Again, we get a classical
radiation field due to the fact that the vacuum fluctuations are washed out
in both limits. These have been amplified to the classical level and we have
produced an intense radiation field. We have a QAMP. 

Higher order corrections can now be computed by the dual Dyson series as usual by the Hamiltonian
\begin{equation}
    H_F = U_F^\dagger(t)\Delta S_z U_F(t)
\end{equation}
that gives us
\begin{eqnarray}
    H_F &=& \sum_n\sum_{S,S',S_x}\langle[n;S',S_x]|\Delta S_z|[n;S,S_x]\rangle|[n;S',S_x]\rangle\langle[n;S,S_x]|
	\\ \nonumber
	&+& \sum_{\stackrel{m,n}{m\neq n}}\sum_{S,S',S_x,S'_x}
	e^{-i(E_{n,S_x}-E_{m,S'_x})t}
	\langle[m;S',S'_x]|\Delta S_z|[n;S,S_x]\rangle|[m;S',S'_x]\rangle\langle[n;S,S_x]|.
\end{eqnarray}
The situation is more involved than the model of a single two-level atom given in sec.\ref{sec6} but
the approach is identical. Here, the main result is that, in the limit $N\rightarrow\infty$, one
gets a classical radiation field. This tends to become an exact result \cite{fra13b}, 
so, even if we started with the strong coupling expansion, we arrived to a non-perturbative result.

\section{Conclusions}
\label{sec8}

We have reviewed the strong coupling expansion as can be applied to time dependent
problems in quantum mechanics. This approach proved to be very fruitful for the
study of a quantum system in different regime, that is, in different regions of the
parameter space of the Hamiltonian.

Having introduced the renormalization group method for removing secularities in
the perturbation series, in a formulation due to Kunihiro, we have built an
algorithm for doing perturbation theory, making very simple the computation
of higher order terms in the series, without any unbounded term in time.

We have seen the method in action by the analysis of a two-level atom in a
single radiation mode. We have obtained the Rabi oscillation in the strong coupling
regime that have been recently observed in Josephson junctions.

The generalization of this model to $N$ two-level atoms
gives a possible description of a new effect
that can be seen as a quantum amplifier (QAMP) of vacuum fluctuations of
the radiation field. The effect appears in the thermodynamic limit
$N\rightarrow\infty$.

We can conclude that a fruitful approach for doing perturbation theory is now
available to analyze quantum systems, in the time domain, in different regions of
the parameter space of the Hamiltonian.

\end{document}